\documentclass[numreferences]{kluwer}

\usepackage{klucite}
\usepackage{graphicx}    

\begin{document}

\baselineskip = 24pt

\begin{opening}
\title{Validating the FLASH Code:  Vortex--Dominated Flows}
\author{Greg Weirs, Vikram Dwarkadas, Tomek Plewa}
\institute{ASCI FLASH Center, University of Chicago
\texttt{vikram@flash.uchicago.edu, tomek@flash.uchicago.edu, weirs@flash.uchicago.edu}}

\author{Chris Tomkins and Mark Marr-Lyon}
\institute{Los Alamos National Laboratory \texttt{ctomkins@lanl.gov, mmarr@lanl.gov}}

\runningauthor{Weirs, Dwarkadas, Plewa, Tomkins, Marr-Lyon}

\begin{abstract}
As a component of the Flash Center's validation program, we compare
FLASH simulation results with experimental results from Los Alamos
National Laboratory.  The flow of interest involves the lateral
interaction between a planar Ma=1.2 shock wave with a cylinder of
gaseous sulfur hexafluoride (SF$_6$) in air, and in particular the
development of primary and secondary instabilities after the passage
of the shock.  While the overall evolution of the flow is comparable
in the simulations and experiments, small-scale features are difficult
to match.  We focus on the sensitivity of numerical results to
simulation parameters.
\end{abstract}

\end{opening}

\section{Introduction}

The impulsive acceleration of a material interface can lead to complex
fluid motions due to the Richtmyer--Meshkov (RM) instability.  Here,
the misalignment of pressure and density gradients deposits vorticity
along the interface, which drives the flow and distorts the interface.
At later times the flow may be receptive to secondary instabilities,
most prominently the Kelvin--Helmholtz instability, which further
increase the flow complexity and may trigger transition to turbulence.

Verification and validation are critical in the development of any
simulation code, without which one can have little confidence that the
code's results are meaningful.  FLASH is a multi--species,
multi--dimensional, parallel, adaptive--mesh--refinement, fluid
dynamics code for applications in astrophysics~\cite{fryols00}.
Calder et al. discuss initial validation tests of the FLASH
code~\cite{calfry02}.  Herein we continue our validation effort by
comparing FLASH simulations to an RM experiment performed at Los
Alamos National Laboratory~\cite{tompre03,zoldi2002}.  

\section{Experimental Facility and Data}

\begin{figure}
\centering
\includegraphics[width=\textwidth,clip=true]{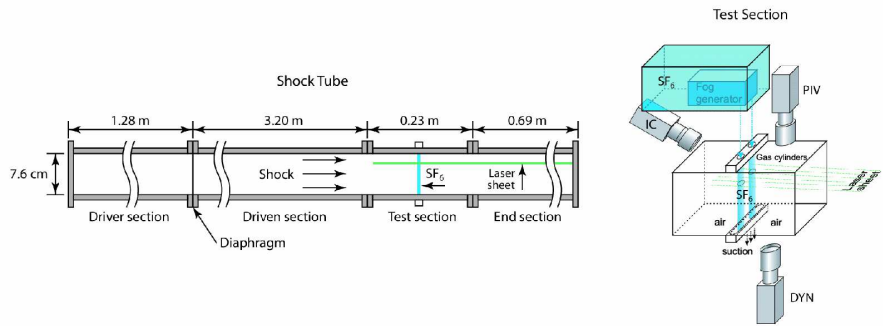}
\caption{LANL shock tube dimensions with test section detail.}
\label{f:shocktube}       
\end{figure}

The experimental apparatus is a shock--tube with a 7.5~cm square
cross--section, as shown in Fig.~\ref{f:shocktube}.  Gaseous SF$_6$
flows from an 8~mm diameter nozzle in the top wall of the shock--tube,
forming a cylinder of dense gas in the otherwise air--filled test
section.  A Ma=1.2 shockwave travels through the shock--tube and
passes through the cylinder. Our interest is in the resulting
evolution of the SF$_6$.  All the experimental data is obtained in a
plane normal to the cylinder axis, 2~cm below the top wall of the test
section.  The experiment is nominally two--dimensional; however, air
diffuses into the SF$_6$ column as it flows downward, thickening the
interface and reducing the peak concentration of the heavy gas.

The initial SF$_6$ distribution (before the shock impact) is
visualized directly by Rayleigh--scattering from the SF$_6$
molecules~\cite{tompre03}.  The pixel intensity in the experimental
image gives only the mole fraction of SF$_6$ relative to the peak mole
fraction, X$_{SF6}$, which must be assumed.  The distribution of
SF$_6$ is only approximately radially symmetric, and the signal is
dominated by noise at the level of about 5--10\%.  Smooth initial
conditions for our simulations are obtained by fitting a
radially--symmetric function to the experimental data.

\begin{figure}
\centering
\includegraphics[height=0.75truein,clip=true]{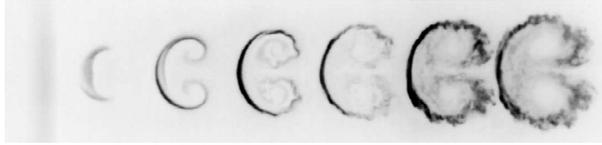}
\caption{Experimental time series of the distribution of SF$_6$.  The
first image corresponds to 50~$\mu$s after shock impact; following
images are at 190, 330, 470, 610, and 750~$\mu$s.  Intensity
corresponds to the mole fraction of SF$_6$.}
\label{f:expt_series}       
\end{figure}

During the experiment the SF$_6$ distribution is indirectly visualized
by visible light scattering off water/glycol droplets, which are
seeded in the SF$_6$.  A sequence of experimental images is shown in
Fig.~\ref{f:expt_series}.  The shock traverses the cylinder in less
than 25~$\mu$s.  The vortex Reynolds number of the flow, as measured
experimentally, is Re$=\Gamma / \nu \approx 5 \times 10^4$, where
$\Gamma$ is the circulation and $\nu$ is the kinematic viscosity.
Each image is taken from a different experimental run.  The
water/glycol droplets can also be used to construct two--dimensional
velocity vectors in the image plane using particle image velocimetry
(PIV)~\cite{prevor00}, but the entire test section must be seeded, so
simultaneous velocity and composition measurements cannot be obtained.

\section{Flowfield Evolution}

As the shock traverses the cylinder, vorticity is deposited along the
interface due to the misalignment of the pressure gradient (normal to
the shock) and the density gradient (normal to the interface.)  The
density gradient arises from the gas composition; SF$_6$ is about five
times as dense as air.  Once the shock has passed through the SF$_6$,
the flow is dominated by a counter--rotating vortex pair, as shown in
Fig.~\ref{f:expt_series}.  Instabilities develop along the distorted
interface at the edge of the primary vortices.  The development and
evolution of the vortex pair and subsequent instabilities at the
interface proceed in a weakly compressible regime.  More precise
descriptions can be found in the
references~\cite{jacobs93,quikar96,zoldi2002}.

\begin{figure}
\centering
\includegraphics[width=\textwidth,clip=true]{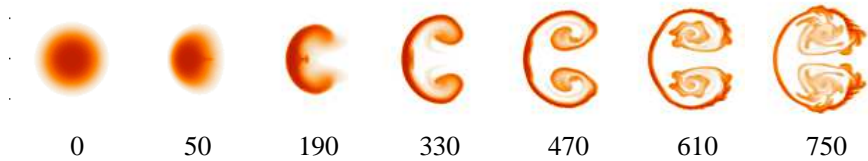}
\caption{Evolution of the SF$_6$, with time elapsed after shock impact
listed in $\mu$s.  The mass fraction of SF$_6$ is shown from a simulation
in which X$_{SF6}$=0.6 and CFL=0.8.}
\label{f:series}       
\end{figure}

The flowfield evolution is driven by flow instabilities and vortex
dynamics, which are sensitive to the initial conditions and noise in
the system.  For validation this sensitivity is desireable because it
provides a severe test for the FLASH code.  Figure~\ref{f:series}
shows a sequence of images from our baseline simulation.  The minimum
grid resolution is 78~microns, the initial peak mole fraction of
SF$_6$ is~0.6, and the Courant (CFL) number is~0.8.  Overall the flow
features in the simulation results are similar to those in the
experimentally obtained images.  Next we describe the effects of
several simulation parameters on the computed results.  The amount and
location of small--scale structure, relative to the experimental data
at 750~$\mu$s, will be used as a qualitative metric.

\section{Results}

In our initial investigations we have focused on the sensitivity of
the computed solutions to several simulation parameters. We have
considered the dependence on the initial maximum mole fraction of
SF$_6$, the mesh resolution, the mesh refinement pattern, and the
Courant number.  We have also compared velocity data, and are
beginning to consider three-dimensional effects.  Here we show results
only for different mesh refinement patterns and Courant numbers.  More
thorough discussion of the results can be found in~\cite{dwaple04}.

Regarding the initial maximum mole fraction of SF$_6$, we find that
simulations where X$_{SF6}$=0.6 seem to match the experimental results
better than when X$_{SF6}$=0.8.  At the higher value, the initial
density gradient is larger; this leads to greater vorticity
deposition, faster instability growth, and consequently, excessive
small scale structure.  However, the time sequences match better for
X$_{SF6}$=0.8.

It is known that unavoidable discretization errors at discontinuous
jumps in grid resolution can act as sources of spurious small--scale
structure~\cite{quirk1991}.  To test this possibility we have run
simulations in which a predetermined area around the vortices is
uniformly refined to the highest resolution.  Compared to fully
adaptive refinement (the default) this approach significantly reduces
the amount of perturbations introduced by the grid adaption process
but increases the computational cost of the simulations.

\begin{figure}
\centering
\includegraphics[height=3.0cm, clip=true]{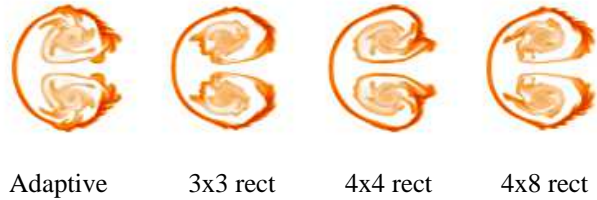}
\caption{Solutions on different grids, 750~$\mu$s after shock impact
at CFL=0.8.  Left to right: fully adaptive grid; $3 \times 3$~cm
refined rectangle; $4 \times 4$~cm refined rectangle; $4 \times 8$~cm
refined rectangle.  In the rightmost image, the refined rectangle
covers the entire spanwise extent of the test section.}
\label{f:rect_ref}       
\end{figure}

In Fig.~\ref{f:rect_ref} we compare the results from a fully adaptive
grid and grids with maximally refined rectangles of $3 \times 3$~cm,
$4 \times 4$~cm, and $4 \times 8$~cm.  The vortex structure is always
less than 2~cm across.  For the different grids the large scale
morphology remains the same, but the shape of the cross--section
visibly differs depending on the grid, as does the amount and location
of small--scale structures.  In particular, differences are noticeable
in the small--scale instabilities present on the vortex rolls.  Since
all other simulation aspects are the same, the differences must
originate with perturbations at jumps in refinement.

\begin{figure}
\centering
\includegraphics[height=3.0cm, clip=true]{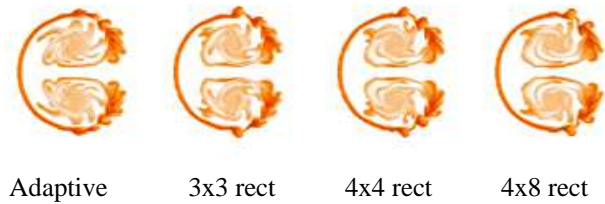}
\caption{Solutions on different grids for CFL=0.2; for other details
see Fig.~\ref{f:rect_ref}.}
\label{f:cfl}       
\end{figure}

We then repeated the simulations on the different grids, but at a
limiting Courant number of CFL=0.2.  The results are shown in
Fig.~\ref{f:cfl}.  We observe much less variation between solutions on
adaptive and locally uniform grids at CFL=0.2 than at CFL=0.8.  One
explanation for these results is that the errors at the fine--coarse
boundaries are larger and lead to stronger perturbations at higher CFL
numbers.  An alternative explanation might be that at higher Courant
numbers, PPM does not adequately compute solutions at these
conditions.  Our simulations indicate that for FLASH, a lower CFL
number leads to more consistent results on different grids.

In addition to the SF$6$ distribution, experimental measurements of
the velocity field in the vicinity of the vortex pair are available.
The particle image velocimetry technique provides two-dimensional
velocity vectors in the image plane.  We find that in the frame of
reference of the vortices, the velocity magnitude and large scale
structure of our simulations match those measured.  The greatest
discrepancies are at smaller scales at the outside edges of the
vortices.

Finally, we executed a speculative three--dimensional simulation.  Our
extension of the initial conditions in the third (cylinder axis)
dimension is purely ad hoc, because we have no corresponding
experimental data.  For this reason this simulation cannot be used as
a validation test for the FLASH code, but we hope it will open a new
line of investigation and discussion.  In the cylinder--axis
dimension, we varied the maximum mole fraction from X$_{SF6}$=0.64 at
the top wall of the test section to X$_{SF6}$=0.47 at the bottom, and
in the same direction the radius of the cylinder increases slightly.
These changes result in greater vorticity deposition at the top wall,
and the vortex pair and instabilities evolve more quickly there.  This
behavior is expected based on the two-dimensional simulations; more
interesting is that the maximum flow velocity in the axial direction
is greater than half that in the spanwise direction by the end of the
simulation (750~$\mu$s).  The stronger vortices have lower core
pressures, and the pressure gradient in the vortex cores accelerates
the flow from the bottom wall toward the top wall.  The air is
preferentially accelerated because of its lower molecular weight and
confinement by the SF$_6$, which acts like the wall of a tube.  For the
initial conditions we have assumed, the axial velocities suggest the
three--dimensional effects are present.

\section{Concluding Remarks}

To date we have made a large number of two--dimensional simulations to
validate the FLASH code for problems dominated by vortex dynamics.  So
far, we have gained a better understanding of the sensitivity of
the computed solutions to simulation parameters such as resolution,
CFL number, and mesh adaption.  While we can recover the overall
morphology, the approximate amount and location of small scale
structure, and velocity field, we must make assumptions (though
reasonable) about the initial conditions to do so.

We continue to work on several fronts.  We lack quantitative,
physically meaningful metrics for comparing simulation and
experimental data.  These metrics are difficult to develop and are
rarely given the attention they deserve.  FLASH simulations do not
currently include a physical model for viscosity, but
resolution--dependent numerical viscosity is present.  Simulations
with a minimum grid spacing of 78$\mu$m exhibit approximately the same
amount of small scale structure as seen in the experimental data,
while results on coarser grids show too little and on finer grids too
much.  We will soon begin simulations with models for physical
diffusion; all the results presented here will then be reviewed.  Our
three--dimensional simulation, despite issues with the initial
conditions, suggests that three--dimensional effects might be
important for this experiment.  We are performing a systematic study
of three--dimensional effects, and hope that experimental data will
become available for comparison.

\vspace*{-0.2in}
\bibliography{hedla04_proc}

\begin{thebibliography}{1}

\bibitem{fryols00}
B.~Fryxell, K.~Olson, P.~Ricker, F.~X. Timmes, M.~Zingale, D.~Q. Lamb,
  P.~Mac{N}eice, R.~Rosner, J.~W. Truran, and H.~Tufo.
\newblock {FLASH}: An adaptive mesh hydrodynamics code for modeling
  astrophysical thermonuclear flashes.
\newblock {\em ApJS}, 131:273--334, November 2000.

\bibitem{calfry02}
A.~C. Calder, B.~Fryxell, T.~Plewa, R.~Rosner, L.~J. Dursi, V.~G. Weirs,
  T.~Dupont, H.~F. Robey, J.~O. Kane, B.~A. Remington, R.~P. Drake, G.~Dimonte,
  M.~Zingale, F.~X. Timmes, K.~Olson, P.~Ricker, P.~Mac{N}eice, and H.~M. Tufo.
\newblock On validating an astrophysical simulation code.
\newblock {\em ApJ Supplement Series}, 143:201--229, November 2002.

\bibitem{tompre03}
C.~Tomkins, K.~Prestridge, P.~Rightley, M.~Marr-Lyon, P.~Vorobieff, and
  R.~Benjamin.
\newblock A quantitive study of the interaction of two
  {R}ichtmyer-{M}eshkov-unstable gas cylinders.
\newblock {\em Physics of Fluids}, 15(4):986--1004, April 2003.

\bibitem{zoldi2002}
C.~A. Zoldi.
\newblock {\em A Numerical and Experimental Study of a Shock-Accelerated Heavy
  Gas Cylinder}.
\newblock PhD thesis, State University of New York at Stony Brook, 2002.

\bibitem{prevor00}
K.~Prestridge, P.~Vorobieff, P.~M. Rightley, and R.~Benjamin.
\newblock Validation of an instability growth model using particle image
  velocimetry measurements.
\newblock {\em Phys.~Rev.~Lett.}, 84(19):4353--4356, May 2000.

\bibitem{jacobs93}
J.~W. Jacobs.
\newblock The dynamics of shock accelerated light and heavy gas cylinders.
\newblock {\em Physics of Fluids A}, 5(9):2239--2247, September 1993.

\bibitem{quikar96}
J.~J. Quirk and S.~Karni.
\newblock On the dynamics of a shock-bubble interaction.
\newblock {\em Journal of Fluid Mechanics}, 318:129--163, 1996.

\bibitem{dwaple04}
V.~Dwarkadas, T.~Plewa, G.~Weirs, C.~Tomkins, and M.~Marr-{L}yon.
\newblock Simulation of vortex--dominated flows using the {FLASH} code.
\newblock In T.~Plewa, T.~Linde, and V.~G. Weirs, editors, {\em Adaptive Mesh
  Refinement -- Theory and Applications}, Springer LNCSE Series.
  Springer-Verlag, 2004.

\bibitem{quirk1991}
J.~J. Quirk.
\newblock {\em An Adaptive Mesh Refinement Algorithm for Computational Shock
  Hydrodynamics}.
\newblock PhD thesis, Cranfield Institute of Technology, UK, 1991.

\end{thebibliography}
\bibliographystyle{unsrt}

\end{document}